\documentstyle[twocolumn,pra,aps,epsfig]{revtex}
\def\be{\begin{equation}}
\def\ee{\end{equation}}
\def\e#1{\label{#1}\end{equation}}
\def\bea{\begin{eqnarray}}
\def\eea{\end{eqnarray}}
\def\ea#1{\label{#1}\end{eqnarray}}
\def\bem#1{\begin{mathletters}\label{#1}}
\def\eml{\end{mathletters}}
\def\r#1{(\ref{#1})}

\begin{document}
\draft
\title{Universal dynamical control of quantum mechanical decay:
Modulation of the coupling to the continuum
}
\author{A. G. Kofman and G. Kurizki}
\address{Department of Chemical Physics, The Weizmann Institute of 
Science, Rehovot 76100, Israel}
%\date{\today}
%\date{July 16, 2001}
\maketitle

\begin{abstract}
We derive and investigate an expression for the dynamically modified 
decay of states coupled to an {\em arbitrary} continuum.
This expression is universally valid for weak temporal perturbations.
The resulting insights can serve as useful recipes for optimized
control of decay and decoherence.
\end{abstract}

The quantum Zeno effect (QZE), namely, the {\em inhibition of the
decay} of an unstable state by its (sufficiently frequent) projective
measurements, has long been considered a basic {\em universal}
feature of quantum systems \cite{kha68}.
Our general analysis \cite{kof00} has revealed the {\em 
inherent impossibility} of the QZE for a broad class of processes, 
including spontaneous emission in open space, as opposed to the 
ubiquitous occurrence of the anti-Zeno effect (AZE), i.e., {\em decay 
acceleration} by frequent projective measurements \cite{lan83}.
Although realistic schemes may well approximate such measurements 
\cite{kof00,coo88,mil88}, there is strong incentive for raising the 
question: 
Are projective measurements the most effective way of modifying the 
decay of an unstable state?
This question is prompted by two important results:
(a) A landmark experiment has demonstrated, for the first time, both
the QZE and AZE by repeated
on-off switching of the coupling between a nearly bound state and the
continuum, using cold atoms that are initially trapped in
an optical-lattice potential \cite{fis01}.
(b) It has been predicted that periodic coherent pulses,
acting between the decaying level and an auxiliary one, can either
inhibit or accelerate the decay into certain model reservoirs 
\cite{aga01}.
In both \cite{fis01} and \cite{aga01}, the repeated interruption of 
the ``natural'' evolution is imperative for decay modification.

In this paper we purport to substantially expand the arsenal of decay
control, whether 
measurement-like (i.e., accompanied by dephasing) or fully coherent.
We derive a universal form of the decay rate of unstable 
states into {\em any} reservoir (continuum),  
modified by weak perturbations with arbitrary time dependence.
The results of Refs. \cite{kof00,lan83,fis01,aga01} are recovered as
limiting cases of this universal form.
Our analysis can serve as a general recipe for {\em optimized} decay 
and decoherence suppression for quantum logic 
operations \cite{bei00} or decay enhancement for the control of chaos 
or chemical reactions \cite{Pre00}.

Consider the decay of a state $|e\rangle$ via its coupling to a 
system, described by the orthonormal basis 
$\{|j\rangle\}$, which forms either a discrete or a continuous 
spectrum (or a mixture thereof).
In its most general form, the total Hamiltonian is 
$\hat{H}_0+\hat{V}(t)+H_1(t)$, where
\be
\hat{H}_0=\hbar\omega_a|e\rangle\langle e|+
\hbar\sum_j\omega_j|j\rangle\langle j|,
\e{5}
with $\hbar\omega_a$ and $\hbar\omega_j$ being the energies of 
$|e\rangle$ and $|j\rangle$, respectively;
\be
\hat{V}(t)=\sum_jV_{ej}(t)|e\rangle\langle j|
+\text{h.c.}, 
\e{6}
denoting the off-diagonal coupling of $|e\rangle$ with the other 
states, which is dynamically modulated, so as to modify the 
{\em static limit} of $\hat{V}$ effecting the natural decay process; 
and
\be
H_1(t)=\hbar\delta_a(t)|e\rangle\langle e|+
\hbar\sum_j\delta_j(t)|j\rangle\langle j|,
\e{7}
standing for the adiabatic (diagonal) time-dependent perturbations
of the energies of the initial 
($|e\rangle$) and final ($|j\rangle$) states, e.g., AC Stark shifts.

\begin{figure}[htb]
\vspace*{-.8cm}
\centerline{\epsfig{file=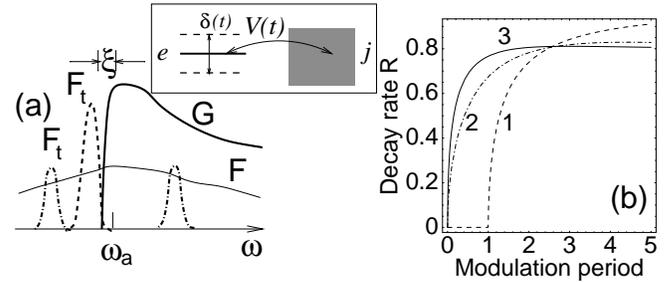,width=8.6cm}}
\vspace*{-.3cm}
\protect\caption{
Decay modification by modulation [Eqs. \protect\r{55}, 
\protect\r{53}].
Inset: Schematic view of the temporal modulation of the shift of level
$e$ and its coupling to a continuum.
(a) $\omega_a$ is near a band edge of
$G(\omega)=C\omega^{1/2}(\omega+\Gamma)^{-1}\theta(\omega)$, where 
$\theta(\omega)$ is the unit step function; then
a small phase shift (dashed peak) is more effective in reducing
the decay rate $R$ than large phase shifts $\phi\simeq\pi$ 
(dash-dotted
peaks) or frequent measurements/random $\epsilon(t)$ (thin curve).
(b) Decay rate $R$ (in units of $R_{\rm GR}$) in 
case (a) with $\omega_a=0.1\Gamma$: reduction by PM [Eq. 
\protect\r{75}] (curve 1 -- $\phi=0.1$, curve 2 -- $\phi=\pi$) and 
frequent impulsive measurements \protect\cite{kof00} (curve 3 -- QZE) 
as a 
function of perturbation period $\tau$ (in units of $\Gamma^{-1}$).
Curve 1 gives the strongest reduction of $R$ at a given $\tau$.
}\label{f1}\end{figure}

We write the wave function of the system, with $|e\rangle$ populated
at $t=0$, as
\bea
|\Psi(t)\rangle=&&\alpha(t)
e^{-i\omega_at-i\int_{0}^t\delta_a(t')dt'}|e\rangle\nonumber\\
&&+\sum_j\beta_j(t)
e^{-i\omega_{j}t-i\int_{0}^t\delta_j(t')dt'}|j\rangle,
\ea{20}
the initial condition being $|\Psi(0)\rangle=|e\rangle$.
Henceforth we treat the generic case wherein the
level shifts and the temporal modulation of $\hat{V}(t)$
are {\em independent} of $j$, i.e., $\delta_j(t)\equiv\delta_f(t)$ and
$V_{je}(t)\equiv\tilde{\epsilon}(t)\mu_{je}$, $\tilde{\epsilon}(t)$ 
being the modulation function (Fig. \ref{f1} -- inset).
Such factorized form of the modulation is commonly valid for weak or
moderate time-dependent fields, which do not appreciably change the 
states of the continuum.
One then obtains from the Schr\"{o}dinger equation that the 
amplitude $\alpha(t)$ obeys 
the {\em exact} integro-differential equation \cite{fed83}
\be
\dot{\alpha}=-\int_{0}^tdt'\epsilon^*(t)\epsilon(t')
\Phi(t-t')e^{i\omega_a(t-t')}\alpha(t').
\e{11}
Here $\Phi(t)=\hbar^{-2}\sum_j|\mu_{ej}|^2e^{-i\omega_j(t)}$ is
the reservoir response (memory) function and the function 
$\epsilon(t)=\tilde{\epsilon}(t)\exp[-i\int_{0}^t\delta_{af}(t')dt']$,
with $\delta_{af}(t)=\delta_a(t)-\delta_f(t)$,
accounts for the modulation of {\em either 
diagonal or off-diagonal elements} of the unperturbed Hamiltonian.

The assumption that the coupling \r{6} is a weak perturbation of 
\r{5} implies that $\alpha(t)$ {\em varies sufficiently slowly} with 
respect to the kernel of Eq. \r{11}, since we then anticipate [cf.
the validity condition \r{76}] decay rates much smaller than 
the rate of change of the reservoir response $\Phi(t)$. 
One can thus make the approximation $\alpha(t')\approx\alpha(t)$ 
on the right-hand side (rhs) of Eq. \r{11}.
Then one can solve Eq. \r{11} and represent the amplitude modulus  
of level $|e\rangle$ in the form
\be
|\alpha(t)|=\exp[-R(t)Q(t)/2],
\e{55}
where we have introduced the fluence
$Q(t)=\int_{0}^t d\tau|\epsilon(\tau)|^2$,
and obtained the decay rate in the {\em universal form}
\be
R(t)=2\pi\int_{-\infty}^\infty d\omega 
G(\omega+\omega_a)F_t(\omega).
\e{53}
Here $G(\omega)=\pi^{-1}\text{Re}\int_0^\infty dte^{i\omega t}\Phi(t)
=\hbar^{-2}\sum_j|\mu_{ej}|^2\delta(\omega-\omega_j)$
is the coupling spectrum, i.e., the density of states weighted by the 
strength of the coupling to the continuum or reservoir;
$F_t(\omega)=|\epsilon_t(\omega)|^2/Q(t)$, with $\epsilon_t(\omega)=
(2\pi)^{-1/2}\int_{0}^t\epsilon(t')e^{i\omega t'}dt'$,
is the (normalized to unity) spectrum of the 
modulation function $\epsilon(t)$ in the ``window'' $(0,t)$.
The result \r{55}, \r{53} is {\em valid to all orders of} $t$, i.e.,
it keeps intact the {\em interferences} between the modulated decay
channels and their non-Markovian effects.
We stress that Eqs. \r{55}, \r{53} apply to the decay of superposed
states $\sum_m\alpha_m|e_m\rangle$ (e.g., in quantum information
schemes), provided all of them decay and are modulated identically.

We now consider some important consequences of the universal form 
\r{55}, \r{53}.
The modulation spectrum $F_t(\omega)$ is roughly characterized by its 
width $\nu_t$ and the frequency shift 
$\Delta_t=\int d\omega\omega F_t(\omega)$.
{\em A modulation may strongly modify the decay rate}
(analogously to the QZE or AZE) whenever 
$\nu_t+|\Delta_t|\agt\xi(\omega_a)$, where 
$\xi(\omega_a)$ {\em is the characteristic spectral interval over
which the weighted density of states $G(\omega)$ changes} near 
$\omega_a$.
In particular, if $\omega_a$ is near the {\em edge of the continuum} 
(as for radiative decay in photonic crystals or vibrational decay in
ion traps,
molecules and solids), then $\xi(\omega_a)$ is the distance between 
$\omega_a$ and the edge \cite{kof00} (Fig. \ref{f1}a).
Only in the opposite limit, $\nu_t+|\Delta_t|\ll\xi(\omega_a)$,
can one approximately set $F_t(\omega)\approx\delta(\omega)$ in Eq. 
\r{53}, yielding $P(t)\approx\exp[-R_{\rm GR}Q(t)]$, where 
$R_{\rm GR}=2\pi G(\omega_a)$ is the {\em extension of the 
Golden-Rule (GR) rate to the case of a time-dependent coupling}.

The modulation function $\epsilon(t)$ can be either random or regular
(coherent) in time, as detailed below.
Consider first the most general coherent {\em amplitude and phase} 
modulation (APM) of the quasiperiodic form,
$\epsilon(t)=\sum_k\epsilon_ke^{-i\omega_kt}$.
Here $\omega_k$ ($k=0,\pm 1,\dots$) are arbitrary discrete 
frequencies with the minimum spectral distance $\Omega$.
For a given function $\epsilon(t)$ one can obtain $-i\omega_k$ and 
$\epsilon_k$ as the poles and residues, respectively, of the Laplace
transform $\hat{\epsilon}(s)$.
If $\epsilon(t)$ is periodic with the period $\Omega$, then 
$\omega_k=k\Omega$, and $\epsilon_k$ become
the Fourier components of $\epsilon(t)$.
For a general quasiperiodic $\epsilon(t)$, one obtains
\be
Q(t)=\epsilon_c^2t+\epsilon_c^2\sum_{k\ne l}\lambda_k\lambda _l^*
\frac{e^{i(\omega_l-\omega_k)t}-1}{i(\omega_l-\omega_k)},
\e{67}
where $\epsilon_c^2=\sum_k|\epsilon_k|^2$ equals the average of 
$|\epsilon(t)|^2$ over a period of the order of $1/\Omega$,
$\lambda_k=\epsilon_k/\epsilon_c$ and
\bea
&&|\epsilon_t(\omega)|^2=\epsilon_c^2t\sum_k|\lambda_k|^2S(\eta_kt/2)
\nonumber\\
&&+\epsilon_c^2\sum_{k\ne l}\lambda_k\lambda_l^*
\frac{1+e^{i(\omega_l-\omega_k)t}-e^{i\eta_kt}-e^{-i\eta_lt}}
{2\pi\eta_k\eta_l}.
\ea{71b}
Here $\eta_k=\omega-\omega_k$, whereas
$S(\eta_kt/2)=2\sin^2(\eta_kt/2)/\pi t\eta_k^2$
is a sinc-function of $\eta_k$ normalized to 1.
 
For $t\gg\Omega^{-1}$ the first term on the rhs of \r{71b} is a sum 
of peaks, whose spacings are much greater than their width $2/t$.
The fast oscillating second term is also peaked at $\omega=\omega_k$, 
but we then find that the ratio of the first to the second terms, and
that of their counterparts in \r{67}, is $\sim(\Omega t)^{-1}\ll 1$.
In the long-time limit, we then neglect these fast oscillating terms
and obtain from Eqs. \r{55}-\r{71b} that 
$P(t)=\exp[-R(t)\epsilon_c^2t]$,
where $R(t)$ in Eq. \r{53} now involves
$F_{t}(\omega)\approx\sum_k|\lambda_k|^2S(\eta_kt/2)$.
For even longer times, exceeding the effective {\em correlation 
(memory) time} of the reservoir,
$t_c\equiv\max_k\{1/\xi(\omega_a+\omega_k)\}$, the functions 
$S(\eta_kt/2)$ become narrower than the respective characteristic 
widths of $G(\omega)$ around $\omega_a+\omega_k$, and one
can set $S(\eta_kt/2)\approx\delta(\eta_k)$.
Then Eq. \r{53} is reduced to
\be
R=2\pi\sum_k|\lambda_k|^2G(\omega_a+\omega_k),
\e{75}
which holds if 
\be
Rt_c\ll 1.
\e{76}
This condition is well satisfied in the regime of interest, i.e., weak
coupling to essentially any reservoir, unless (for some $k$)
$\omega_a+\omega_k$ is extremely close to a sharp feature in 
$G(\omega)$, e.g., a band edge \cite{kof94}.
Hence, the long-time limit of the general decay rate \r{53} under the
APM is a sum of the GR rates,
corresponding to the resonant frequencies shifted by $\omega_k$, with
the weights $|\lambda_k|^2$.
Formula \r{75} provides a {\em simple general recipe} for 
manipulating the decay rate by APM.
Its powerful generality allows for the {\em optimized} control of
decay, not just for a single level, but also for {\em band}
characterized by a spectral distribution $P(\omega_a)$ (e.g.,
inhomogeneous or vibrational spectrum).
We can then choose $\lambda_k$ and $\omega_k$ in Eq. \r{75} so as to
minimize the decay of \r{55} convoluted with $P(\omega_a)$.
The following limits of \r{75} will be now analyzed.

(i) {\em Monochromatic perturbation}: 
Let $\epsilon(t)=\epsilon_0e^{-i\delta_{af}t}$. 
Then $R=2\pi G(\omega_a+\Delta)$, where
$\Delta=\delta_{af}=\text{const}$ is an AC Stark shift.
In principle, such a shift may drastically enhance or 
suppress $R$ relative to $R_{\rm GR}$.
It provides the {\em maximal variation} of $R$ achievable with an 
external perturbation, since it does not involve any averaging 
(smoothing) of $G(\omega)$ incurred by the width of $F_t(\omega)$:
the modified $R$ can even {\em vanish}, if the shifted frequency 
$\omega_a+\Delta$ is beyond the cutoff frequency of the coupling, 
where $G(\omega)=0$ (Fig. \ref{f1}a,b).
Conversely, the increase of $R$ due to a shift can be
much greater than that achievable with the AZE \cite{kof00}.
In practice, however, AC Stark shifts are usually small for (CW)
monochromatic perturbations, whence pulsed perturbations should often 
be used, resulting in multiple $\omega_k$ shifts according to \r{75}.

(ii) {\em Impulsive phase modulation (PM)}:
Let the phase of the coupling amplitude jump by an amount $\phi$ at 
times $\tau,2\tau,\dots$.
Such modulation can be achieved by a train of identical, equidistant,
narrow pulses of nonresonant radiation, which produce pulsed AC Stark
shifts $\delta_{af}(t)$ in \r{7}.
Now $\epsilon(t)=e^{i[t/\tau]\phi}$, where $[\dots]$ is the integer 
part.
One then obtains that $\epsilon_c=1$ and $Q(t)=t$.
The decay is given by Eqs. \r{55} and \r{53}, where $F_t(\omega)$ can
be obtained in a closed form.
For sufficiently long times one can use Eq. \r{75}.
The poles and residues of 
$\hat{\epsilon}(s)=(1-e^{-s\tau})/[s(1-e^{i\phi-s\tau})]$ yield
$\omega_k=2k\pi/\tau-\phi/\tau$ and
$|\lambda_k|^2=4\sin^2(\phi/2)/(2k\pi-\phi)^2$.
For {\em small phase shifts}, $\phi\ll 1$, the $k=0$ peak dominates, 
$|\lambda_0|^2\approx 1-\phi^2/12$, whereas
$|\lambda_k|^2\approx\phi^2/4\pi^2k^2$ for $k\ne 0$.
In this case one can retain only the $k=0$ term in Eq. \r{75} [unless
$G(\omega)$ is changing very fast].
Then the modulation acts as a constant shift $\Delta=-\phi/\tau$. 
With the increase of $|\phi|$, the difference between the $k=0$ and 
$k=1$ peak heights diminishes, {\em vanishing} for $\phi=\pm\pi$.
Then $|\lambda_0|^2=|\lambda_1|^2=4/\pi^2$, i.e., $F_t(\omega)$ for 
$\phi=\pm\pi$ contains {\em two identical peaks symmetrically shifted 
in opposite directions} (Fig. \ref{f1}a) [the other peaks
$|\lambda_k|^2$ decrease with $k$ as $(2k-1)^{-2}$, totaling 0.19].

The above features allow one to adjust the modulation parameters for a
given scenario to obtain an optimal decrease or increase of $R$.
The PM scheme with a small $\phi$ is preferable near the continuum
edge (Fig. \ref{f1}a,b), since it yields a spectral shift in the 
required direction (positive or negative).
The adverse effect of $k\ne 0$ peaks in $F_t(\omega)$ then scales as 
$\phi^2$ and hence can be significantly reduced by decreasing 
$|\phi|$.
On the other hand, if $\omega_a$ is near a {\em symmetric} peak of
$G(\omega)$, $R$ is reduced more effectively for $\phi\simeq\pi$, as
in Ref. \cite{aga01}, since the main peaks of $F_t(\omega)$ at 
$\omega_0$ and $\omega_1$ then shift stronger with $\tau^{-1}$ than 
the peak at $\omega_0=-\phi/\tau$ for $\phi\ll 1$.

(iii) {\em Amplitude modulation (AM)} of the coupling arises, e.g.,
for radiative-decay modulation due to atomic motion through a 
high-$Q$ cavity or a photonic crystal \cite{she92} or for atomic 
tunneling in optical lattices with
time-varying lattice acceleration \cite{fis01,niu98}.
Let the coupling be
turned on and off periodically, for the time $\tau_1$ and
$\tau_0-\tau_1$, respectively, i.e., $\epsilon=1$ for 
$n\tau_0<t<n\tau_0+\tau_1$ and $\epsilon=0$ for 
$n\tau_0+\tau_1<t<(n+1)\tau_0$ ($n=0,1,\dots$).
Now $Q(t)$ in \r{55} is the total time during which the coupling is 
switched on.
This case is also covered by Eq. \r{75}, where 
the parameters are now found to be $\epsilon_c^2=\tau_1/\tau_0$,
$\omega_k=2k\pi/\tau_0$, $|\lambda_0|^2=\tau_1/\tau_0$,
$|\lambda_k|^2=(\tau_1/\tau_0)\text{sinc}^2(k\pi\tau_1/\tau_0)$
($k\ne 0$).

\begin{figure}[htb]
\vspace*{-5cm}
\hspace{-1.cm}
\centerline{\epsfig{file=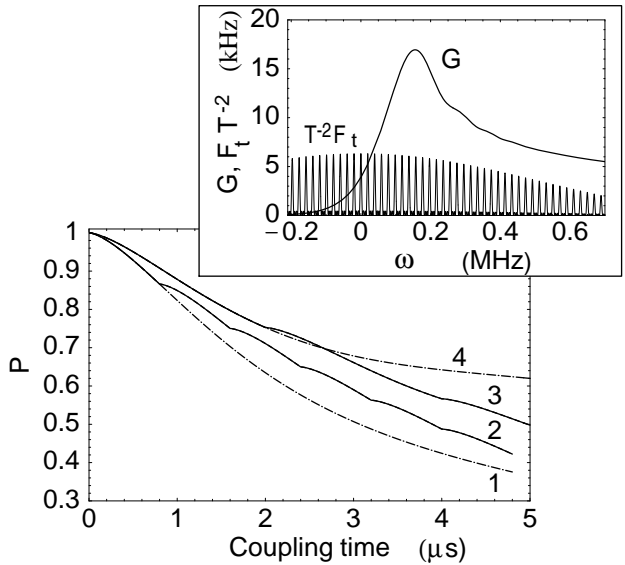,width=5.in}}
\vspace{-5cm}
\protect\caption{
Tunneling of sodium atoms in optical lattices perturbed by AM scheme:
the decay probability $P(t)$ as a function of the total coupling time.
Curves 1, 4 -- decay without modulation.
Curve 2 -- QZE (decay slowdown compared to 
curve 1) for $\tau_1=0.8\ \mu$s, $\tau_0=50.8\ \mu$s.
Curve 3 -- AZE (decay speedup compared to 
curve 4) for $\tau_1=2\ \mu$s, $\tau_0=52\ \mu$s.
Inset: 
The coupling spectrum $G(\omega+\omega_a)$ and the scaled modulation 
function $T^{-2}F_{4\tau_0}(\omega)$ for the conditions of curve 2.
Here $T=\omega_gd/(\pi a)$, where $a=15$ km/s$^2$ is the lattice 
acceleration and $d=295$ nm is the lattice period.
$\omega_g=91$ kHz, $\omega_g T=2.05$ (for curves 1, 2, 5);
$\omega_g=116$ kHz, $\omega_g T=3.32$ (for curves 3, 4).
}
\label{f2}
\end{figure}

It is instructive to consider the limit wherein $\tau_1\ll\tau_0$ and 
$\tau_0$ is much greater than the correlation time of the continuum,
i.e., $G(\omega)$ does not change significantly 
over the spectral intervals $(2\pi k/\tau_0,2\pi(k+1)/\tau_0)$.
In this case one can approximate the sum \r{75} by the integral 
\r{53} with $F_t(\omega)\approx(\tau_1/2\pi)
\text{sinc}^2(\omega\tau_1/2)$, characterized by the spectral 
broadening $\sim 1/\tau_1$ (Fig. \ref{f2} -- inset).
Then Eq. \r{53} for $R$ reduces to that obtained 
when ideal projective measurements are performed at intervals 
$\tau_1$ \cite{kof00}.

Thus the AM scheme {\em can imitate measurement-induced (dephasing) 
effects} on quantum dynamics, if the interruption intervals $\tau_0$
exceed the {\em correlation time of the continuum}.
This indeed has been observed \cite{fis01} for atom tunneling in 
optical lattices whose tilt (acceleration) was periodically modulated 
as above.
For its analysis we have used the approximate expression for 
$\Phi(t)$ 
obtained in \cite{niu98}, which yields the reservoir spectrum 
$G(\omega+\omega_a)$ (Fig. \ref{f2} -- inset), with one maximum at 
$\omega\sim\omega_g$, $\hbar\omega_g$ being the lattice band gap.
The decay probability $P(t)$, calculated in Fig. \ref{f2} (curves 1-4)
for parameters similar to \cite{fis01}, {\em completely 
coincides} with that obtained for ideal impulsive 
measurements at intervals $\tau_1$ \cite{kof00} and
demonstrates either the QZE (curve 2) or the AZE (curve 3) behavior.

The universal Eq. \r{53}, which is a result of {\em unitary} 
analysis, is valid also when $\epsilon(t)$ 
is a {\em stationary random process}.
If such a process is characterized by the correlation time $\nu^{-1}$,
one can use a master equation to show that, for $t\gg\nu^{-1}$,
we have $P(t)\approx e^{-Rt}$, where the decay rate (provided that
$R\ll\nu$) still has the general form \r{53}, but with
\be
F_t(\omega)\rightarrow F(\omega)=\pi^{-1}\epsilon_c^{-2}$Re$
\int_0^\infty\overline{\epsilon^*(t)\epsilon(0)}e^{i\omega t}dt,
\e{85}
$F(\omega)$ being the normalized spectrum of the random process and 
$\epsilon_c^2=\overline{|\epsilon(t)|^2}$, where the overbar denotes 
ensemble averaging.
Expression \r{53} with the substitution \r{85} is {\em completely 
analogous} to the universal formula describing {\em measurement 
effects} on quantum evolution in \cite{kof00}.
This analogy between unitary and measurement effects stems from the
ability to {\em emulate} projective measurements by the dephasing of
the level evolution caused by classical random fields 
\cite{kof00,mil88}.
 
There may, however, be a notable difference between projections and
random-field dephasing.
Projective measurements at an effective rate $\nu$, whether impulsive 
or continuous, usually result in a broadened (to a  width $\nu$)
modulation function 
$F(\omega)$, without a shift of its center of gravity, 
$\Delta=\int d\omega\omega F(\omega)\approx 0$ \cite{kof00,coo88}.
This feature was shown \cite{kof00} to be responsible for
either the standard QZE scaling, $R\sim 1/\nu$, or the AZE scaling.
In contrast, a weak and broadband chaotic field such that 
$|\chi|\overline{I}\ll\nu_B$, where $\overline{I}$ is the mean 
intensity, $\nu_B$ is the bandwidth, and $\chi$ is the effective 
polarizability, would give rise to a
Lorentzian dephasing function $F(\omega)$ in \r{85} with a 
substantial shift $\Delta=\chi\overline{I}$.
This shift would have much stronger effect on $R$ than the QZE or AZE 
caused by the width $\nu\sim\chi^2\overline{I}^2/\nu_B\ll|\Delta|$. 

We have presented here a general theory of dynamically modulated
quantum decay, which offers new insights into the possibilities of
controlling its non-Markovian dynamics by off-resonant electromagnetic
fields.
Its unified form \r{55}, \r{53} encompasses, as special cases, all 
the modulation schemes of current interest, satisfying the 
factorization condition [cf. Eq. \r{11}] \cite{fis01,aga01,aga00}.
Whereas its limit \r{85} may imitate measurement effects 
(the QZE and AZE), the modulation or spectral-shift parameters allow 
us to ``engineer'' (suppress or enhance)
more effectively the decay into a given reservoir.
Thus, measurements are shown to have {\em no advantage} as a means of 
either suppressing or enhancing decay compared to APM.
Moreover, the coherent nature of APM makes it much more appropriate
than measurements for decoherence suppression in quantum information
applications, which require {\em reversible}
transformations of quantum superposed states.

This work has been supported by EU (ATESIT Network), the US-Israel 
BSF, and the Ministry of Absorption (A. K.).

\end{document}